\documentclass[twocolumn]{aastex631}

\newcommand{\unit}[1]{\ \mathrm{#1}}
\newcommand{\solarrad}{\unit{R}_{\odot}}
\newcommand{\solarmass}{\unit{M}_{\odot}}
\newcommand{\ten}[1]{\times 10^{#1}}

\newcommand{\mbh}{M_{\mathrm{bh}}}
\newcommand{\ms}{M_{*}}

\newcommand{\kb}{k_{\mathrm{b}}}

\newcommand{\lbol}{L_{\mathrm{bol}}}
\newcommand{\tbb}{T_{\mathrm{bb}}}
\newcommand{\rbb}{R_{\mathrm{bb}}}

\newcommand{\mh}{m_{\mathrm{H}}}
\usepackage{overpic}

\begin{document}

\title[Optical Bound TDEs]{Optical Appearance of Eccentric Tidal Disruption Events}

\author[0000-0001-7984-9477]{Fangyi (Fitz) Hu}
\affiliation{School of Physics and Astronomy, Monash University, Clayton VIC 3800, Australia}
\affiliation{OzGrav: The ARC Centre of Excellence for Gravitational Wave Discovery, Australia}

\author[0000-0002-4716-4235]{Daniel J. Price}
\affiliation{School of Physics and Astronomy, Monash University, Clayton VIC 3800, Australia}

\author[0000-0002-6134-8946]{Ilya Mandel}
\affiliation{School of Physics and Astronomy, Monash University, Clayton VIC 3800, Australia}
\affiliation{OzGrav: The ARC Centre of Excellence for Gravitational Wave Discovery, Australia}



\begin{abstract}

Stars approaching supermassive black holes can be tidally disrupted. Despite being expected to emit X-rays, TDEs have been largely observed in optical bands, which is poorly understood. In this Letter, we simulate the tidal disruption of a $1~M_\odot$ main sequence star on an eccentric ($e=0.95$) orbit with a periapsis distance one or five times smaller than the tidal radius ($\beta = 1$ or $5$) using general relativistic smoothed particle hydrodynamics. We follow the simulation for up to a year post-disruption. We show that accretion disks in eccentric TDEs are masked by unbound material outflowing at $\sim10,000~$km/s. Assuming electron scattering opacity, this material would be visible as a $\sim100~$au photosphere at $\sim10^4~$K, in line with observations of candidate TDEs.



\end{abstract}



\section{Introduction} \label{sec:intro}

When a star approaches a supermassive black hole (SMBH), it will be disrupted by the gravitational tidal force, resulting in a tidal disruption event (TDE; \citealt{Rees1988}). An accretion disk is formed from the stellar debris \citep{Hayasaki2013, Bonnerot2016a, Hayasaki2016, Steinberg2022, Ryu2023a}. A small fraction of the star's mass ($\lesssim 10 \%$) \citep{Metzger2016} would be accreted by the SMBH, which releases a large amount of energy sufficient to expand the rest of the gas and create observable radiation.

X-rays were originally expected to be observed from TDE accretion disks, but most TDEs are actually detected in optical/UV \citep[e.g.][]{Rees1988,Gezari2006,Gezari2008,van-Velzen2011,Holoien2019,van-Velzen2021a,Leloudas2022} instead of X-ray \citep[e.g.][]{Auchettl2017,Gezari2017,Sazonov2021a,Malyali2023a}. Optical spectra show broad spectral lines consistent with 10,000-20,000 km/s outflows \citep[e.g.][]{van-Velzen2011}. Several possible sources of the optical emission have been proposed, including stream-stream collisions \citep[e.g.][]{Piran2015}, disk winds \citep{Miller2015} and the re-emission of lower-energy photons from a reprocessing layer that down-converts X-rays emitted at deeper locations \citep[e.g.][]{Loeb1997,Metzger2016}.

The reprocessing layer model has been verified by \citet{Guillochon2014}, as least in the case of PS1-10jh, where a lightcurve consistent with observations was produced by a model that accounts for both the accretion disk and a reprocessing layer. Understanding the origin of such a reprocessing layer may therefore be key to relating simulations to observations.


Early simulations of TDEs focused on trying to understand the mass accretion rate, $\dot{M}$, onto the black hole and its dependence on stellar structure \citep[e.g.][]{Evans1989, Lodato2009, Guillochon2013}. These studies already showed that typically $\dot{M}$ would exceed the Eddington rate for 1--2 years following the initial disruption. Subsequent studies have tried to use the assumed $\dot{M}$  to predict the lightcurve during the super-Eddington phase using semi-analytic models \citep[e.g.][]{Strubbe2009a, Strubbe2011, Lodato2011, Mockler2019}, or by assuming a pre-existing disk in simulations \citep[e.g][]{Shen2014, Curd2019}. More recent studies have successfully simulated disk formation initiated by stream-stream collisions caused by relativistic precession \citep{Hayasaki2013, Bonnerot2016a, Hayasaki2016, Andalman2022, Steinberg2022, Ryu2023a}, but simulations were followed for a limited time following the disruption, used polytropic stellar models, Newtonian or post-Newtonian approximations to gravity, and no attempt was made to produce synthetic lightcurves (with the notable exception of \citet{Steinberg2022} who computed synthetic lightcurves for up to 65 days from their post-Newtonian simulations of parabolic TDEs).


In this Letter, we study disk formation in an eccentric TDE, similar to the previous study by \citet{Bonnerot2016a} but with four key differences: i) we use real stellar profiles rather than polytropes; ii) we evolved the simulations over a much longer timescale; iii) we perform our simulations in full general relativistic hydrodynamics in the Schwarzschild metric rather than approximate potentials and most importantly iv) we use the optical depth of the surrounding material to predict the optical appearance, rather than relying on the mass fallback rate as a proxy for luminosity. The latter turns out to provide a relatively straightforward explanation for the otherwise mysterious nature of optical emission from TDEs in the form of the long-hypothesised reprocessing layer.

\section{Methods} \label{sec:methods}

We simulate the tidal disruption of a $\ms = 1 \solarmass$ solar-type star around a $\mbh = 10^{6} \solarmass$ non-spinning (Schwarzschild) SMBH on an orbit with eccentricity $e = 0.95$ and considering two penetration factors $\beta \equiv R_{\rm tidal}/ R_{\rm peri} = 5$ and $1$ \citep{Rees1988}, using the general relativistic smoothed particle hydrodynamics (GRSPH) code \textsc{Phantom} \citep{Price2018,Liptai2019a}. For our chosen star we have $R_{\rm tidal} \equiv R_* (M_{\rm BH}/ M_*)^{1/3} = 100 R_\odot \approx 47 GM_{\rm BH}/c^2$ giving a pericentre (apocentre) distance for $\beta=5$ of $20 R_\odot$ ($780 R_\odot$) with a semi-major axis of 400$R_\odot$ and hence an orbital period of 0.93 days (22 hours) for $\beta=5$ and $10.4$ days for $\beta = 1$. For reference the Schwarzschild radius of a $10^6$  M$_\odot$ black hole is $4.2 R_\odot$. We place the star initially at apocentre, meaning the first pericentre passage occurs at 0.46 days.

\begin{figure*}
        \includegraphics[width=2\columnwidth]{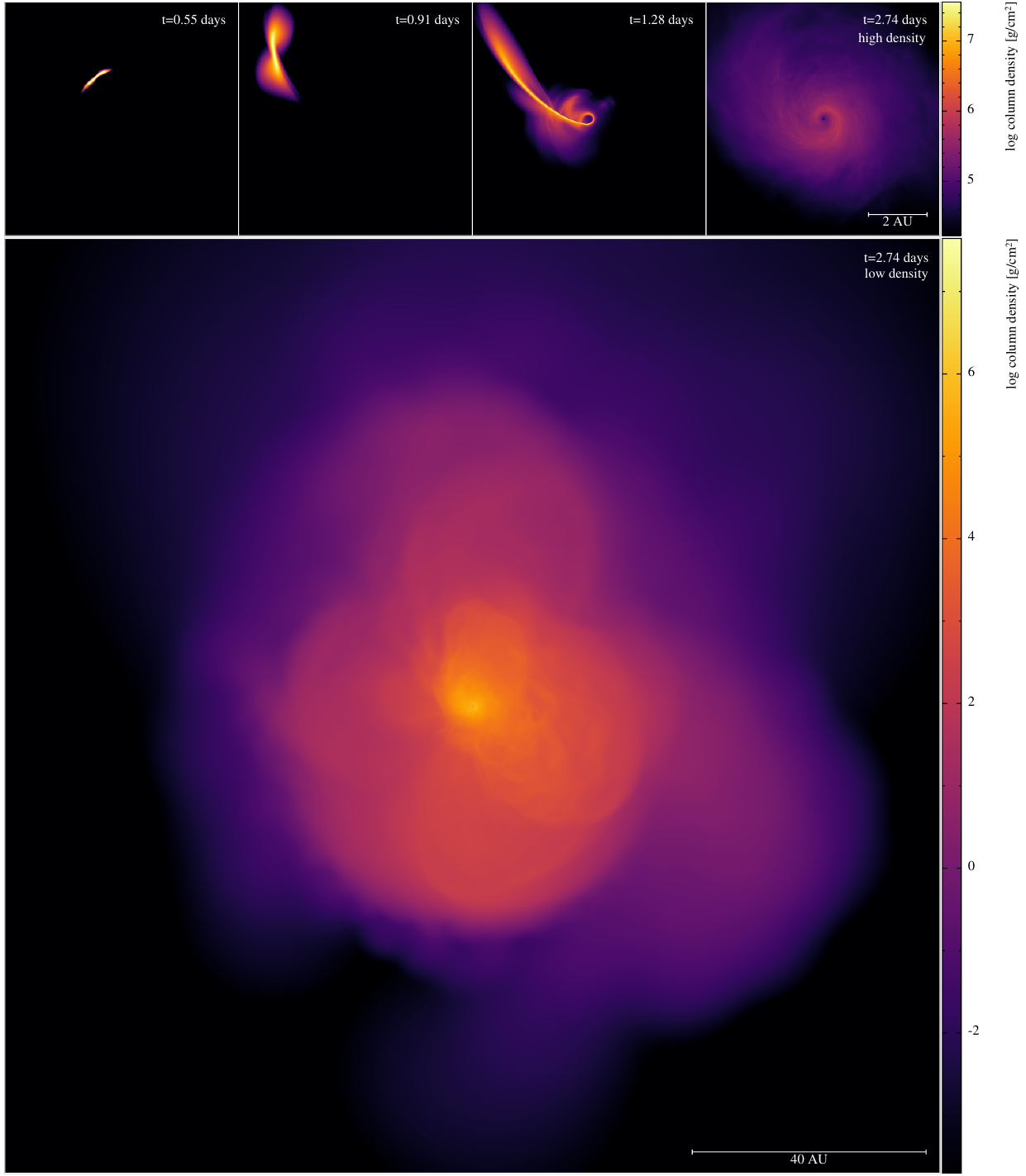}
    \caption{Column density during disk formation of the simulated TDE viewed from $z=\infty$, normal to the initial stellar orbital plane. The upper panels illustrate the regions with high column density, i.e. $\rho_{\min} = 2\ten{4} \unit{g} \unit{cm}^{-2}$ and the lower panel includes larger regions with lower column density, i.e. $\rho_{\min} = 2\ten{-4} \unit{g} \unit{cm}^{-2}$. Time is since first apoapsis, with first periapsis at 0.46 days given the 0.93 day period. The high-density disk has radius of $\sim 2 \unit{au}$ which is consistent with \citet{Bonnerot2016a}. The full TDE, even at this early stage, extends to radius of $\sim 40 \unit{au}$ and the low-density region is consistent with the reprocessing layer model which buries the disk.}
    \label{fig:x y no opac}
\end{figure*}


\subsection{Initial Star}
Starting with a $1 \solarmass$ pre main sequence star with solar metallicity of $0.0134$ \citep{Asplund2009}, we evolve the star with the stellar evolution code \textsc{MESA} \citep{Paxton2011,Paxton2013,Paxton2015,Paxton2018,Paxton2019} until the current solar age, i.e. $4.6 \unit{Gyr}$. The star has a central density of 180 g/cm$^3$, which is 21 times higher than the central density of a $\gamma = 5/3, 1 \solarrad, 1 \solarmass$ polytrope.

\citet{Guillochon2013} investigated the effect of stellar structure of TDEs but did not go as far as a real star. The main difference we find when simulating the tidal disruption of real stars is that for $\beta = 1$ it takes 5 orbits to completely disrupt the star whereas on the same orbit a polytropic star would be disrupted in one passage. For the deeper $\beta = 5$ encounter, the star is disrupted on the first passage regardless of the stellar profile. 

We relax the star into hydrostatic equilibrium prior to placing it in orbit, following the {\sc relax-o-matic} procedure outlined in Appendix~C of \citet{Lau2022}. That is, we relax the star with the temperature profile fixed, until the ratio of kinetic-to-potential energy is $\lesssim 10^{-7}$. For the GR code we perform the relaxation in the Minkowski metric, adding the self-gravity of the star via a metric perturbation of the form $h_{\mu \nu} = {\rm diag}[-2\Phi/c^2,0,0,0]$. 

\subsection{Simulation parameters}
We employ an adiabatic equation of state (EoS) $P = (\gamma - 1)\rho u$ with $\gamma = 5/3$ throughout the simulation (see below), where $P$, $\rho$ and $u$ are the pressure, density and specific internal energy respectively. Since we use the Kerr metric in Boyer-Lindquist coordinates, we set an accretion boundary within the last stable orbit at $5GM_{\rm BH}/c^2$, where particles are removed from the simulation. We use the default shock-capturing parameters in the GRSPH code, with $\alpha_{\rm AV} = 1$, $\beta_{\rm AV} = 2$ and shock conductivity $\alpha_{\rm cond} = 0.1$ and evolve the specific entropy $s$ as the conserved variable \citep[see][]{Liptai2019a}. We employ the default cubic spline kernel \citep{Monaghan1985} with a ratio of smoothing length to particle spacing of $1.2$, corresponding to a mean neighbour number of $\sim 60$ in 3D.



\subsection{Computing the temperature}
Our equation of state assumes a gas-pressure-dominated adiabatic EoS, but the post-pericenter material should be radiation-pressure-dominated, corresponding to $\gamma = 4/3$ instead of $5/3$. For this paper we do not include radiation pressure in the simulation itself, and in preliminary comparisons with simulations that include radiation pressure we find only small differences.

However, we find that it is critical to account for the contribution from radiation pressure when interpreting the internal energy in the simulation in terms of temperature, which determines the opacity and optical appearance. Specifically, we compute the temperature ${T}$ from $\rho$ and $u$ by solving the following quartic equation for $T$:
\begin{equation} \label{eq:gas+rad eos}
   \frac32 \frac{\kb T}{\mu \mh} + \frac{{a T}^{4}}{\rho} = u,
\end{equation}
where $\mu=0.5988$ is the mean molecular weight taken from our initial stellar model. Since the temperature itself is not employed in the hydrodynamical calculation, this was done in post-processing.

\subsection{Obtaining lightcurves from simulations} \label{sec:optical}
Once the simulations are complete, we estimated the optical appearance of the accretion flow in our simulations by ray tracing over a square grid of pixels in the image plane ($x_i$,$y_j$), assuming parallel rays reaching the observer at $z=\infty$, by solving the equation of radiative transfer and propagating rays through the column density projection of each SPH particle. For all the results shown in this Letter, we use an image plane parallel to the initial stellar orbital plane and $z$ is the normal direction. As the radiation passes through particle $n$, the total intensity changes as
\begin{equation} \label{eq:ray tracing}
    I_{\nu,n} = I_{\nu, n-1} e^{-\tau_{n}} + S_{\nu, n} (T_{\mathrm{p},n}) \left( 1 - e^{-\tau_{n}} \right),
\end{equation}
where $I_{\nu, n-1}$ is the total intensity before reaching the particle and $S_{\nu}$ is the source function. The optical depth through each particle is computed according to
\begin{equation} \label{eq:optical depth}
    \tau_n =  \kappa m_{n} Y(|x_i - x_n|,|y_j - y_n|, h_n),
\end{equation}
where $m_n$ is the SPH particle mass, $h$ is the smoothing length, $\kappa$ is the opacity, and $Y$ is the column-integrated kernel function with dimensions of inverse area (i.e. the 3D spherical kernel function integrated through z, see Eq. 29 of \citealt{Price2007}).

A key problem in TDEs is that the accretion flow should be highly scattering-dominated \citep{Leloudas2022}. The source function with scattering is
\begin{equation}
    S_{\nu}(T) = (1 - \epsilon_{\nu}) B_{\nu}(T) + \epsilon_{\nu} J_{\nu},
\end{equation}
where $\epsilon_{\nu}$ is the albedo, $B_{\nu}$ is the blackbody function and $J_{\nu}$ is the mean intensity at the point. In the environment of the TDE reprocessing layer, where atoms are all ionised, electron scattering is expected to dominate, i.e. $\epsilon_{\nu} \approx 1$. However, with abundant scatterings, photons are promptly thermalised into local equilibrium, giving $J_{\nu} \approx B_{\nu}(T_{\mathrm{local}})$. 
We can therefore approximate each SPH particle as a local blackbody emitter, i.e. $S_{\nu}(T) \approx B_{\nu}(T)$, and use the scattering opacity $\kappa = \kappa_{\rm es}$ since we assume scattering would dominate the extinction. We compute the Thomson electron scattering opacity according to
\begin{equation} \label{eq:thomson opacity}
    \kappa_{\mathrm{es}} = \sigma_{\mathrm{e}} n_{\mathrm{e}} / \rho,
\end{equation}
where $\sigma_{\mathrm{e}}$ is the Thomson electron scattering cross section and $n_{\mathrm{e}}$ is the electron density. The electron density is determined from the ionisation fraction $x_{\mathrm{H}}$ which is calculated by solving the Saha equation assuming pure hydrogen gas, i.e. 
\begin{equation} \label{eq:saha}
    \frac{x_{\mathrm{H}}^{2}}{1-x_{\mathrm{H}}} = \frac{1}{n_{\mathrm{H}}} \left( \frac{2\pi m_{\mathrm{e}} \kb T_{\mathrm{p}}}{h^{2}} \right)^{3/2} e^{-\chi_{\rm H}/\kb T_{\mathrm{p}}},
\end{equation}
where $n_{\mathrm{H}}, m_{\mathrm{e}}, h$ and $\chi_{\rm H}$ are the hydrogen number density, the electron mass, the Planck constant and the ionisation energy of hydrogen respectively. The number density is thus $n_{\rm e} = \chi_{\rm H}n_{\rm H} = \chi_{\rm H}\rho/m_{\rm H}$. 

While the above simplified treatment allows us to compute an approximate optical appearance, we caution that a full solution of the radiative transfer problem in scattering-dominated flows is needed to carefully test the assumptions above, and that our method likely provides an upper limit for the true photosphere radius.




\begin{figure*}

\begin{overpic}[width=2\columnwidth]{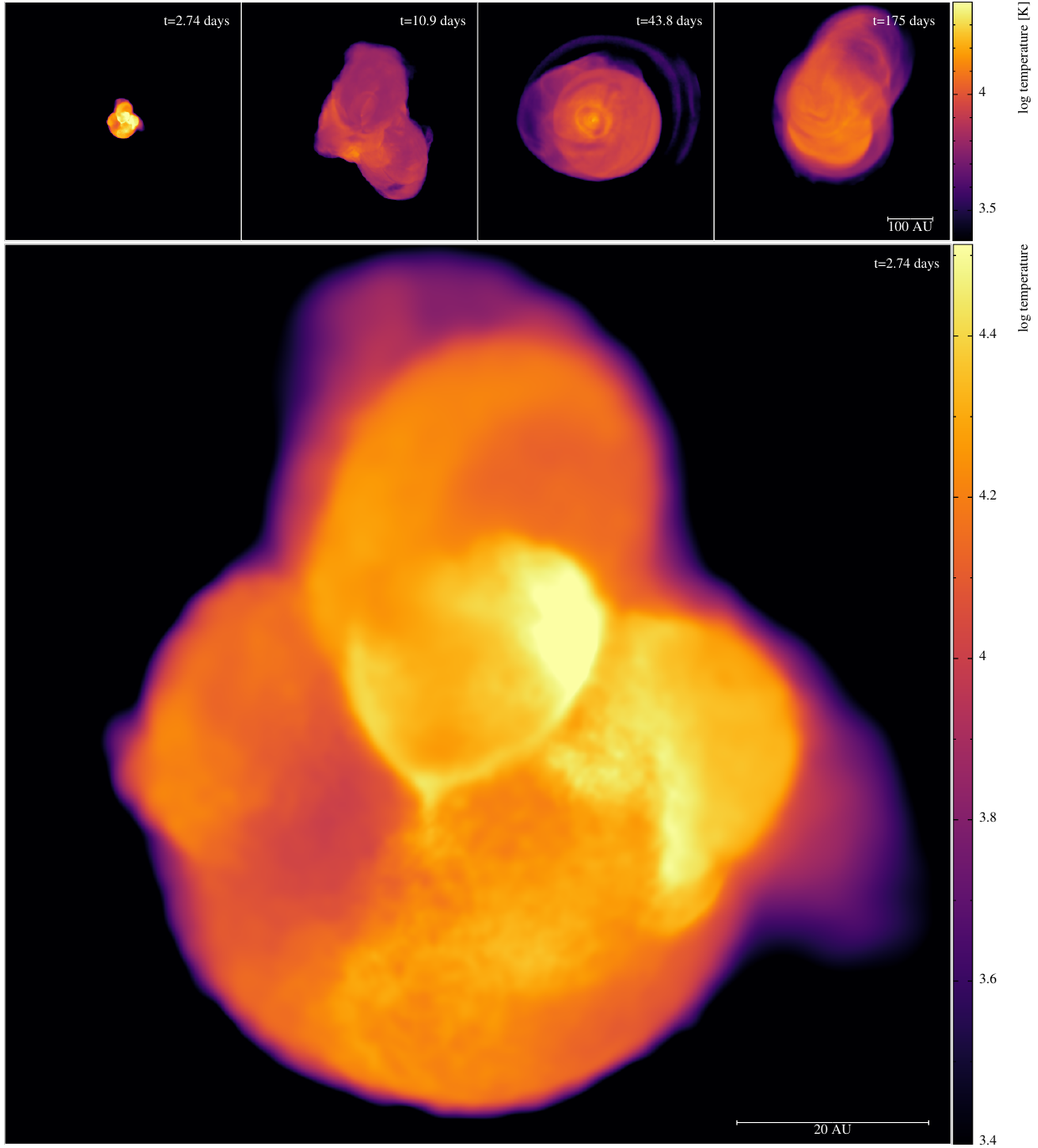}
 \put (88.6,79) {\scriptsize\rotatebox{90}{[K]}}
\end{overpic}
    \caption{Temperature at photosphere of the simulated TDE, viewed with electron scattering opacity and with the observer at $z=\infty$, normal to the initial stellar orbital plane (as in Figure~\ref{fig:x y no opac}). In the first and second top panels ($t \lesssim 20~$days), the photosphere is expanding asymmetrically, while in the third and fourth panels ($t \gtrsim 20~$days), the photosphere remains roughly constant at $\sim 100 \unit{au}$ due to the expansion of the reprocessing layer and the contraction of the photosphere, consistent with observations \citep[e.g.][]{van-Velzen2021a}.}
    \label{fig:opac}
\end{figure*}

\subsubsection{Lightcurve}\label{sec:lightcurve}
From our procedure above  we compute the flux by integrating the intensity at each wavelength over the image plane
\begin{equation}
F_\nu = \int I_\nu {\rm d}x {\rm d}y.
\end{equation}
With the specific flux $F_{\nu}$, we can plot the spectrum as $\nu F_{\nu}$.
Following the procedure in real observations, we then fit a blackbody to our spectrum considering only the ZTF optical observing band of 367.6 nm to 901.0 nm \citep{Bellm2019,Rodrigo2020}, considered as a box bandpass for simplicity. The blackbody temperature $\tbb$ and radius $\rbb$ are obtained from the fit and the corresponding bolometric luminosity is obtained according to
\begin{equation} \label{eq:bolometric luminosity}
    \lbol = 4 \pi \rbb^{2} \sigma \tbb^{4},
\end{equation}
where $\sigma$ is the Stefan-Boltzmann constant. 

\section{Results}

\subsection{Disk formation}
\label{sec:disk} 

Figure~\ref{fig:x y no opac} shows four snapshots of the column density, $\Sigma$, of our $\beta=5$ simulation, showing the debris stream generated by the first pericentre passage (top left two panels). At $t \approx 1$ day (third panel), after the second pericenter passage of the leading debris stream, the stream self-collides, leading to prompt disk formation after approximately 3 orbits (third and fourth panel). 

The upper panels of Figure~\ref{fig:x y no opac} show the material at high column density ($1.6\times10^{4}$ to $3.2\times10^{7} \unit{g} \unit{cm}^{-2}$), illustrating the process of disk formation. Disk formation can be seen to occur rapidly (within 2 days) due to the relativistic apsidal advance and subsequent self-intersection of the debris stream (third panel). At 2.74 days, there is $0.44 \solarmass$ and $0.40 \solarmass$ of material in rotationally-dominated ($|v_{\phi}| > |v_{r}|$) and radially-dominated (($|v_{\phi}| < |v_{r}|$) motion respectively, whereas the remaining $0.16 \solarmass$ is accreted. At 100 days, only $0.05 \solarmass$ remains in the rotationally-dominated disk/torus and $0.64 \solarmass$ is outflowing.

\subsection{Optical Appearance}

The bottom panel of Figure~\ref{fig:x y no opac} shows the same snapshot as the fourth upper panel but zoomed out and shown over a much larger range in column density. This figure shows i) that there is a large amount of low-density material expelled to large distances ii) that there is no clear boundary between the accretion disk and the surrounding outflow and iii) this material would likely be optically thick since the optical depth to electron scattering $\tau \approx \kappa_{\rm es} \Sigma$ exceeds unity for $\Sigma \gtrsim 5~$g/cm$^2$ assuming $\kappa_{\rm es} = 0.2~$cm$^2$/g, corresponding to $\log \Sigma \sim~1$ on the colour bar. Thus, everything appearing orange in the plot would be optically thick.


It is commonly speculated that a reprocessing layer is responsible for the optical radiation seen in observations. Such a reprocessing layer is speculated to be the low-density region surrounding the accretion disk. In Figure~\ref{fig:x y no opac}, we see that the low-density gas is ejected from the center asymmetrically which forms the reprocessing layer. The reprocessing layer covers the accretion disk in every direction so it could block and reprocess the X-rays as desired.


\begin{figure}
    \centering
    \includegraphics[width=\columnwidth]{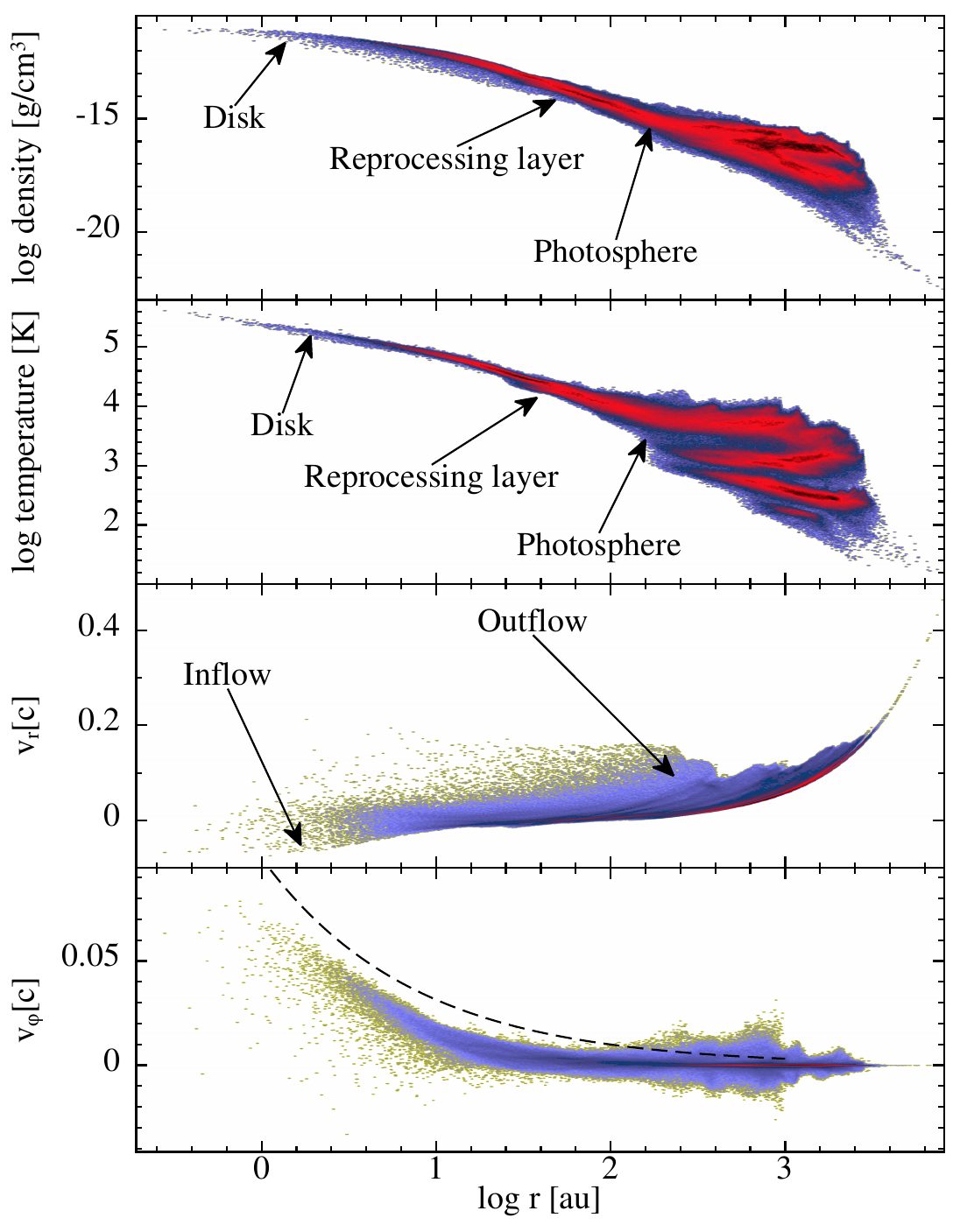}
    \caption{Profiles of the density (first), temperature (second), radial (third) and azimuthal (fourth) velocities at $t=100$ days. Colors in the plot represent the density of particles in the parameter space (yellow-blue-red-black corresponds to low to high). Dashed line shows Keplerian orbital velocity. The material is approximately spherically symmetric inside the estimated photosphere and more asymmetrical outside. A small amount of material inflows within $r<1~$au, with the remaining material outflowing.}
    \label{fig:tde profile}
\end{figure}

Figure~\ref{fig:opac} illustrates this in more detail, showing the later evolution of the material shown in Figure~\ref{fig:x y no opac} but using a simple ray trace with the electron scattering opacity, showing the temperature of material that would be observed at the last scattering surface.


The photosphere keeps expanding asymmetrically with the reprocessing layer in the early stage (first and second upper panels), when the temperature is $\gtrsim 10^{4} \unit{K}$. As the outer temperature drops below $\sim 10^{4} \unit{K}$ with the expansion, ionised hydrogen recombines, opacity drops, and the photosphere contracts. Due to the counterbalance between the expansion of the reprocessing layer and the contraction of the photosphere, the photosphere radius remains roughly constant at $\sim 100 \unit{au}$ (third and forth upper panels in Figure~\ref{fig:opac}). This is consistent with blackbody radii of $\sim~10$--$100~\unit{au}$ inferred from optical observations of TDEs \citep[e.g.][]{van-Velzen2021a}.


\subsubsection{Properties of the reprocessing layer}
Figure \ref{fig:tde profile} shows a snapshot of the density, temperature, radial and azimuthal velocity profiles of the TDE gas envelope after 100 days. Most of the gas, at this stage, is outflowing. The radial velocity generally increases with radius, with typical speeds of $\sim $10--30,000 km/s (0.03--0.1 c) with the maximum being higher than $0.4 \unit{c}$. For $r \gtrsim 100$ au we find material in homologous expansion with $v \propto r$. Only a small portion of gas within radius $\lesssim 1~$au flows towards the BH and might be accreted. Material at $r < 10~$au can be seen to orbit at sub-Keplerian speeds, indicating a pressure-dominated thick disk or torus (with aspect ratio $H/R \sim 0.75$).

The density and temperature decrease with radius with typical temperatures of $\sim10^4$ K at the approximate location of the photosphere. Inside $r \sim 100$~au material is approximately spherically symmetric, while for $r \gtrsim 100$~au material is distributed more asymmetrically --- seen by the larger range in temperature and density at the same radius. For $r \lesssim 10~$au we find $T \propto r^{-0.5}$ and $\rho \propto r^{-1}$ while for $r=10$--100 au we find $T \propto r^{-0.9}$ and $\rho \propto r^{-2}$.

\begin{figure}
    \centering
    \includegraphics[width=0.95\columnwidth]{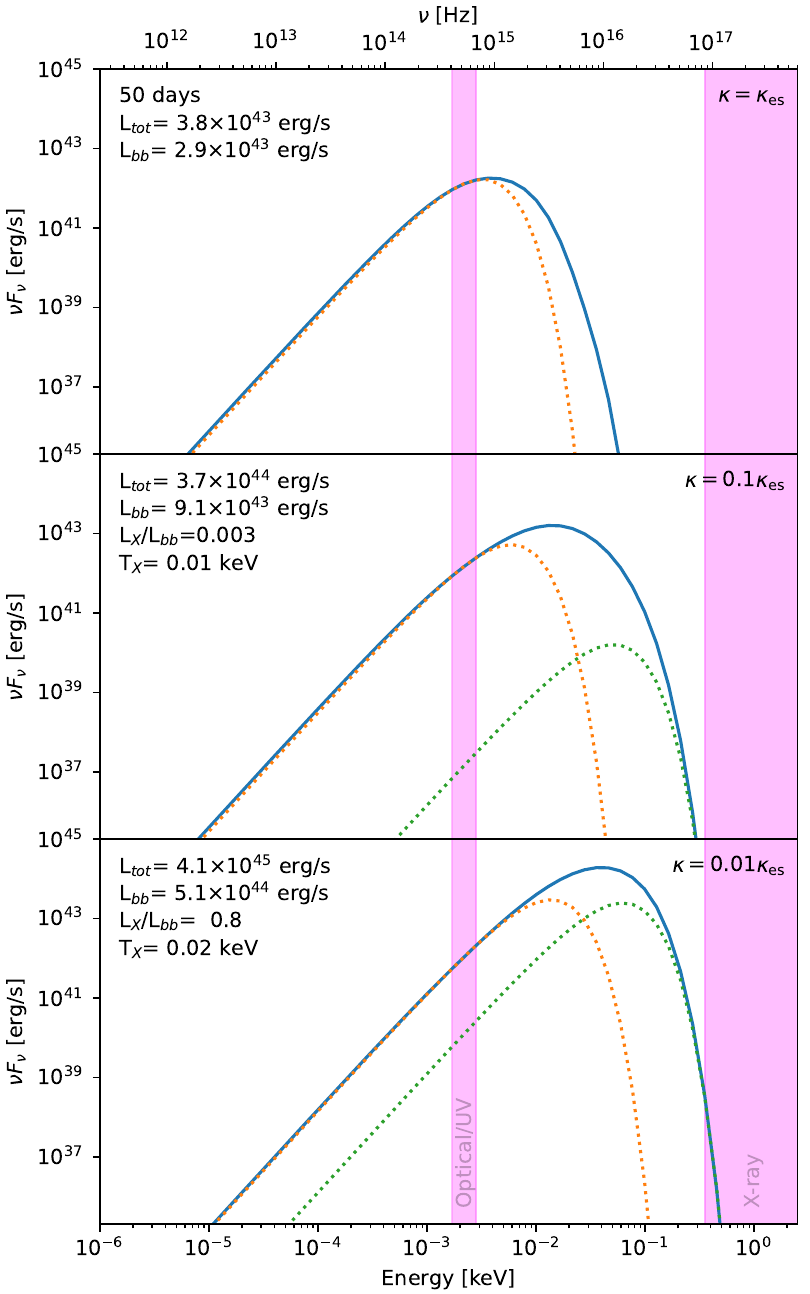}
    \caption{Example spectra (blue) obtained by our ray tracing procedure, for an observer at $z=\infty$, perpendicular to the orbital plane, for our $\beta=5$ simulation. Blackbodies are fitted to the ZTF optical/UV (orange) and Swift X-ray (green) band (magenta boxes) separately. Different opacities are used in each panel, i.e. $\kappa = \kappa_{\rm es}$, $0.1\kappa_{\rm es}$ and $0.01\kappa_{\rm es}$ in the top, middle and bottom panel respectively. The lower the opacity, the deeper we can look into the gas envelope and the harder the spectrum.}
    \label{fig:spectrum}
\end{figure}

\subsection{Lightcurve}
Figure~\ref{fig:spectrum} shows the synthetic spectra (blue curve) obtained with the simplified ray tracing procedure described in Section~\ref{sec:optical}, shown at $t=50$ days, using the electron scattering opacity (top panel), an opacity reduced by a factor of 10 (middle panel) and by 100 (lower panel). The versions with lowered opacity are shown to capture our uncertainty in the true location of the photosphere given our simplified radiative transfer model (see discussion).

The spectrum in all cases is broadly consistent with those found in observations of candidate TDEs, with a rising flux across the optical band with a peak at $\nu \sim 10^{15}$--$10^{16}$ Hz \citep[e.g.][]{van-Velzen2011}. The orange dotted line in each panel shows the corresponding blackbody fit to the optical/UV band, whereas the green dotted line shows the corresponding fit to the Swift X-ray band. The lower panel shows that X-ray emission is possible as an extension of the thermal spectrum if one is able to see deeper. A single blackbody cannot describe the whole spectrum in any cases as anticipated by \citet{van-Velzen2011}.

\begin{figure}
	\includegraphics[width=\columnwidth]{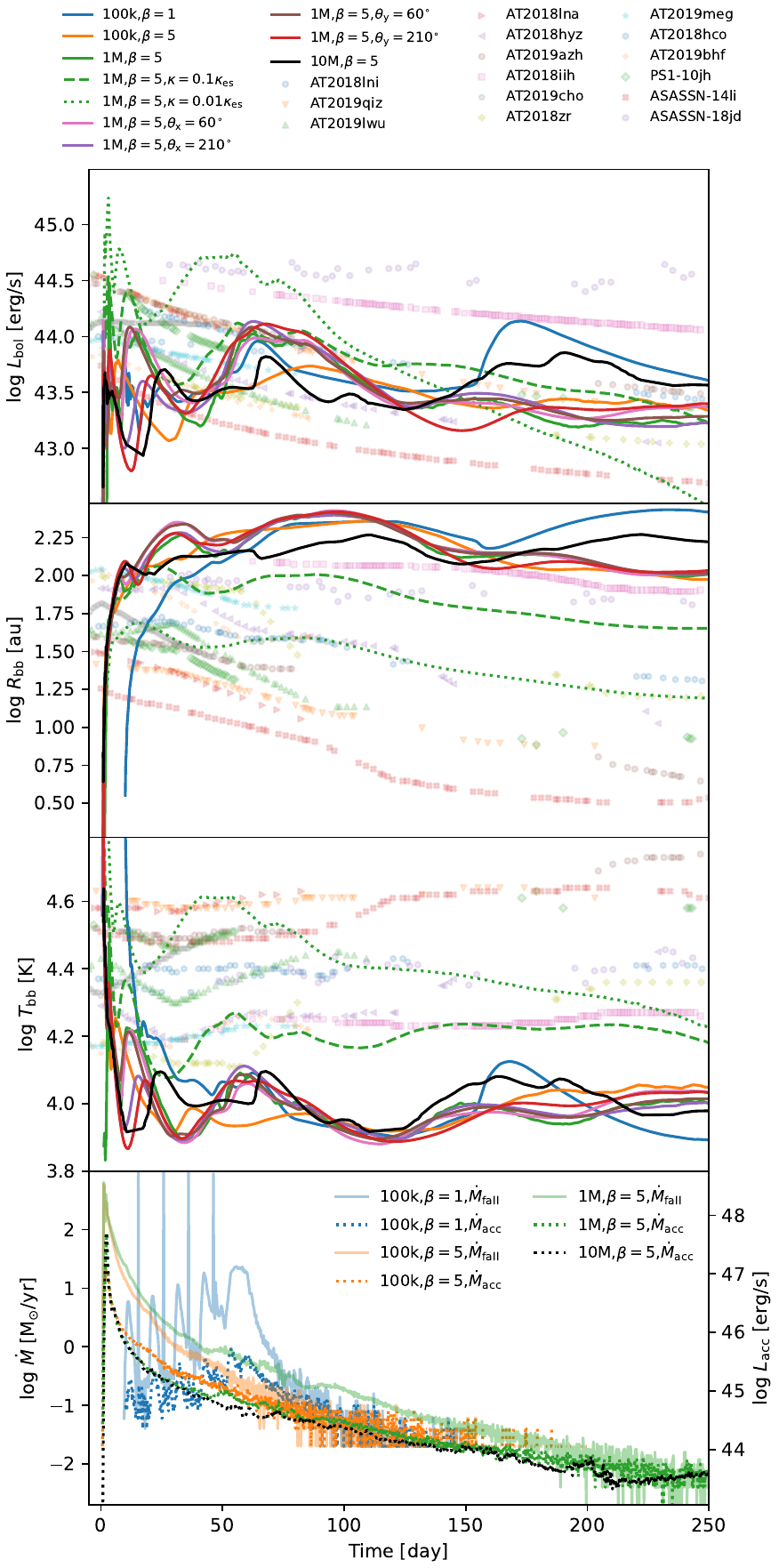}
    \caption{Bolometric luminosity (first), blackbody radius (second), blackbody temperature (third) and mass fallback and accretion rates (fourth) of the TDE simulation. Both the luminosity and radius increase rapidly at the beginning and then remain relatively constant, whereas the temperature drops sharply and stays roughly constant. Compared to observations (coloured dots), the temperature and radius are systematically lower and higher respectively. The luminosity follows neither the fallback rate nor the accretion rate.}
    \label{fig:lightcurve}
\end{figure}


Figure~\ref{fig:lightcurve} shows the resulting lightcurves, including the bolometric luminosity (first panel), blackbody radius (second panel) and temperature (third panel) obtained from the optical blackbody fits. $\lbol, \rbb$ and $\tbb$ of the simulations are $\sim 10^{43.5} \unit{erg} \unit{s}^{-1}$, $100 \unit{au}~(\approx 1.5\times10^{15} \unit{cm})$ and $10^{4} \unit{K}$ respectively. We compare our lightcurves to the data points (coloured dots) showing observed lightcurves from recent optically-detected TDEs taken from \citet{van-Velzen2021a}.

Our inferred luminosities are consistent with the observations, being in the range of $10^{43}-10^{45}$ erg/s. Assuming a photosphere at the electron scattering surface (solid lines) gives blackbody radii ($\sim 100$ au) and temperatures ($\sim 10^4$ K) within the right order of magnitude of the observations, but systematically higher and lower than the observations, respectively. Our model, therefore, is realistic but misses certain physics (see Section~\ref{sec:discussion}). 

There are more fluctuations in our synthetic lightcurve than seen in the observations. Comparing simulations at different resolutions (black compared to green and orange lines) shows that these fluctuations are resolution-dependent and therefore untrustworthy, although the values of temperature, luminosity and radius around which these fluctuations occur do not show an obvious trend with resolution. Surprisingly, simulations at a lower penetration factor ($\beta = 1$; blue line) or seen from different viewing angles (rotated around x- or y-axis by $\theta_{\rm x}$ or $\theta_{\rm y}=60^{\circ}$ or $210^{\circ}$; pink, purple, brown and red lines) also produce lightcurves essentially indistinguishable to those with $\beta = 5$, suggesting that the outflow mechanism is generic as long as prompt circularisation of the debris occurs. However, this may be exacerbated by the lack of photospheric cooling in our model.

The dashed and dotted green lines in Figure~\ref{fig:lightcurve} show the lightcurves obtained with the opacity dropped by factors of 10 and 100, respectively. Of all the effects we investigated, our assumed opacity has the largest effect on the resultant blackbody temperature and radius, giving higher temperatures and smaller radii for reduced opacities, more in line with the observations. This can be understood from Figure~\ref{fig:tde profile}, since seeing deeper into the reprocessing layer gives higher temperatures and smaller sizes.

The fourth panel of Figure~\ref{fig:lightcurve} shows the mass fallback rate $\dot{M}_{\rm fall}$ (solid lines) and accretion rate $\dot{M}_{\rm acc}$ (dotted lines) of each simulation. We defined fallback and accreted material as material inside of $150G\mbh/c^{2}~\approx~1.5~$au \citep{Cufari2022a} and $5G\mbh/c^{2} \approx 0.05~$au which is inside the innermost stable circular orbit, respectively. Once a particle enters the corresponding radii, it is added to the total fallback or accreted mass, respectively, from which we computed gradients by smoothing on a timescale of $\approx 4$ hours. A particle could return to the fallback radius several times before final accretion or ejection. Every peak in $\dot{M}_{\rm fall}$ corresponds to a peak in $\dot{M}_{\rm acc}$ at approximately the same time regardless of the penetration factor. Multiple peaks are observable for $\beta=1$ which corresponds to each pericenter passage before full disruption (the first peak is missed due to passage during being shorter than the time interval between saved dumps). The peaks in the lightcurve occur at $\sim 1~$day after the final peaks in $\dot{M}_{\rm acc}$ which are consistent with the advection timescale from the innermost stable circular orbit to the photosphere. $\dot{M}_{\rm fall}$ is $\sim 1$ order of magnitude higher than $\dot{M}_{\rm acc}$ before and at the peak time but the two rates gradually converge afterward. Assuming radiative efficiency equivalent to 10\% of the accreted mass-energy, we plot a secondary y-axis ($L_{\rm acc} = 0.1m_{\rm acc}c^2$) in the fourth panel. The nominal accretion luminosity $L_{\rm acc}$ would exceed the Eddington limit ($\sim 10^{44}~$erg/s for a $10^{6} \solarmass$ BH) by several orders of magnitude for $t \lesssim 100$ days, indicating that most of the accretion energy is either advected or goes into powering the kinetic outflow rather than being emitted as radiation.

\section{Discussion} \label{sec:discussion}

We have presented two simulations of the tidal disruption of a $1\solarmass$ star by a supermassive black hole. Our simulations improve on previous simulations of disk formation in eccentric tidal disruption events \citep[e.g.][]{Hayasaki2013, Hayasaki2016, Bonnerot2016a,Andalman2022, Steinberg2022, Ryu2023a} by using both full GR and a main sequence star instead of a polytrope. While we make fewer approximations than previous authors, the circularisation process is similar, driven by relativistic apsidal advance of the debris stream as it passes close to the black hole. Our circularisation timescale is slower than \citet{Bonnerot2016a}, i.e. $\sim 2$ days compared to 22 hr, due to the more centrally concentrated stellar structure. The most important difference in our investigation is that we evolved our simulations for up to one year following the onset of accretion. Equally important is that we considered the effects of the low density material that is ejected from close to the black hole (large panel in Figure~\ref{fig:x y no opac}), whereas they only showed plots of the high column density material from their simulations (equivalent to our top row of panels in Figure~\ref{fig:x y no opac}). 

Considering electron scattering opacity, we predict that this ejected material is optically thick ($\tau_{\rm es} \gg 10^5$; see Figure~\ref{fig:x y no opac}), and thus we found that the accreting torus is buried inside a reprocessing layer that is remarkably similar to that predicted by phenomenological models of TDE outflows \citep{Loeb1997, Bogdanovic2004, Strubbe2009a, Guillochon2014, Coughlin2014, Jiang2016, Metzger2016, Mockler2019}. For example, \citet{Loeb1997} suggested `a steady, spherical, optically thick envelope around the black hole' which they termed the `Eddington envelope', while \citet{Metzger2016} suggested that the non-accreted material should form an outflow with typical speeds of $\sim 10^4$ km/s. \citet{Guillochon2014} and \citet{Mockler2019} developed this idea further, showing that one can successfully fit TDE lightcurves with a reprocessing layer in the form of an outflow powered by the accretion flow. The outflows we find in the simulations are consistent with these models, and also with observations of broad emission lines indicating outflows with speeds of 5,000-20,000 km/s \citep{van-Velzen2011, Gezari2012a, Gezari2015, Nicholl2019}.



We found that the magnitude of the predicted luminosity is (mostly) independent of the numerical resolution, and surprisingly also insensitive to the penetration factor, implying that the outflow structure has universal properties once accretion onto the black hole commences. Our predicted photospheric temperatures and blackbody radii of $\sim 10^4$ K and $\sim 100$ au are of the correct order of magnitude, although systematically lower and higher, respectively, compared to observations. We found that the predicted temperature and fitted blackbody radius depend most strongly on the assumed depth of the photosphere, which we have varied in our model by artificially lowering the opacity in our simplified ray-tracing scheme (dotted and dashed green lines in Figure~\ref{fig:lightcurve}). Our predicted photospheric temperatures for $t \gtrsim 50$ days are around the hydrogen recombination temperature ($\sim 10^{4}~$K) which makes sense, since the recombination temperature sets the location of the photosphere. The observed temperature, however, is higher, indicating photons might be scattered up from hotter, significantly ionized regions --- or that a single-temperature thermal distribution is an even worse fit to the integrated lightcurve than the models suggest.  Determining the true optical emission from a simulation would require post-processing with full radiative transfer in a highly scattering-dominated regime \citep{Leloudas2022}. In particular one should consider the separation between the thermalisation surface (where photons are given their temperature) and the photosphere (last scattering surface) \citep[see e.g.][]{Masuda2015}.



Although the observations also show a range in temperature, luminosity and radius evolution, the shape of our lightcurves are by no means a perfect match to the observations. This could be due to the eccentric nature of our simulated TDE, our approximate radiative transfer, or our use of an adiabatic equation of state which allows no energy to escape from the photosphere. The latter, however, is unlikely to have a significant effect on the evolution of our TDE. The total energy radiated in all bands is $\sim 10^{51}~$erg, which is less than 2\% of the total available energy ($\gtrsim 5\times10^{52}~$erg; computed from the difference in the total thermal + kinetic energy at $t=1$ year compared to the start of the simulation). The lack of radiation pressure could be responsible as it is more sensitive to temperature than gas pressure, i.e. $\propto T^{4}$ compared to $T$, so the reprocessing layer would respond to changes more rapidly and have a smoother temperature profile and thus produce a smoother lightcurve. Preliminary simulations that we have performed that include radiation pressure confirm this to be the case.

We observe no X-rays from the current simulation until $t\approx 1$yr if we assume that the photosphere corresponds to the last scattering surface for electrons. The maximum temperature has cooled down to $\sim 10^{4} \unit{K}$ at the end of a year, so its thermal emission peaks in optical/UV band. No observable X-rays can be emitted through thermal emission at this stage. 
At earlier times, for example 50 days (Figure~\ref{fig:spectrum}) to 100 days post-disruption (Figure~\ref{fig:tde profile}), the maximum temperature is $\gtrsim 10^{5} \unit{K}$ which peaks in  soft X-ray. If the core thermal emission can escape during this stage, X-ray TDEs might be observed. The escape probability, however, would be sensitive to the opacity of the reprocessing layer. By manually lowering the opacity by a factor of 10 or 100, we see X-ray escapes at $t<100~$days with luminosity of $\sim 10^{42}$ and $10^{44}~{\rm erg\ g}^{-1}$ respectively. Although the reprocessing layer is not spherically symmetric, surprisingly we found the viewing angle is not critical in observing X-rays. The X-ray luminosity is similar in all directions with slight offsets in the brightening and decaying rate. \citet{Gezari2017} have, on the other hand, observed X-ray brightening and UV fading in the TDE ASASSN-15oi at $\sim 300$ days post-disruption, when our simulated TDE cannot produce X-rays. ASASSN-15oi is likely to be a parabolic or extremely eccentric TDE that has continuous material fallback and accretion to heat the core region.

The source of energy that powers the TDE outflow is also a mystery. Possible sources include accretion \citep[e.g.][]{Hayasaki2013}, stream-stream collision \citep[e.g.][]{Jiang2016} and nozzle shocks \citep[e.g.][]{Ryu2023a}. In our simulations, on the second fallback and stream-stream collision ($t=1.28$ day), the total energy (kinetic and thermal) is $\sim 10^{20}, 6\times10^{18}$ and $6\times10^{19}~{\rm erg\ g}^{-1}$, with only $1\%$ to $10\%$ being thermal energy, at $r=0.06, 1$ and $0.1~{\rm au}$ respectively which correspond to the innermost stable circular orbit (accretion radius), stream-stream collision radius and pericenter (nozzle shock radius). Most of the energy comes from the vicinity of the black hole in the form of kinetic energy which implies accretion, instead of nozzle shocks or shocks from the stream-stream collisions, as the ultimate source of energy for the TDE outflow.

Early investigations of TDEs commonly assumed the luminosity follows the mass accretion rate and predict lightcurves from the accretion rate \citep[e.g.][]{Evans1989,Lodato2009,Strubbe2009a,Strubbe2011,Lodato2011}. Our simulations (Figure~\ref{fig:lightcurve}) suggest this may not be true for eccentric TDEs after the peak. Most of the energy is released before the peak ($<2~$days for $\beta=5$ and $<60~$days for $\beta=1$), but trapped to expand the gas envelope and gradually radiated over time. The one-day offset in luminosity peak and accretion rate peak indicates an optically thick gas envelope with radius $\sim 100~$au has already formed at the time of the peak. The lightcurve therefore traces the accretion rate before the peak but becomes independent afterward. However, \citet{Mockler2019} suggested that a lightcurve that more closely tracks the fallback rate may provide a better match to observations.






\section{Conclusions}
We have presented GRSPH simulations of the tidal disruption of a $1 \solarmass$ star on a bound, eccentric orbit $e = 0.95$ and penetration factor $\beta = 5$ or 1 around a $10^{6} \solarmass$ Schwarzschild SMBH. Our main findings are:
\begin{enumerate}
\item Prompt accretion commences on a timescale of several days, consistent with previous studies.
\item The thick accretion disk is buried in a large amount of low-density material outflowing with speeds of $\sim 10^4$ km/s. This material is consistent with the long-hypothesised reprocessing layer needed to explain optical emission from TDEs. 
\item Considering electron scattering opacity, we produced synthetic lightcurves that can be directly compared to optical lightcurves from observed candidate TDEs.
\item We find photospheric temperatures and blackbody radii of $\tbb\sim 10^4~$K and $\rbb\sim 100~ $au, respectively, broadly consistent with observed TDEs.

\item With no cooling of the photosphere and assuming electron scattering opacity, our model tends to under-predict the temperatures and over-predict the photosphere radii, and does not allow for X-ray emission. A faster-decaying lightcurve,  more consistent with the observed decay, can be reproduced by lowering the opacity in synthetic observations of the model. This argues for the importance of incorporating accurate radiation transport in future studies.
\end{enumerate}

\begin{acknowledgments}
We acknowledge the CPU time on OzSTAR and Ngarrgu Tindebeek funded by the Victorian and Australian governments and Swinburne University.


\end{acknowledgments}

\bibliography{ref}{}
\bibliographystyle{aasjournal}



\end{document}